\documentclass[aps,pra,amsmath,twocolumn]{revtex4}

\usepackage{amsmath,amssymb}
\usepackage{graphicx}
\usepackage{color}
\usepackage{xspace}

\newcommand{\bra}[1]{\langle #1|}
\newcommand{\ket}[1]{|#1\rangle}
\newcommand{\braket}[2]{\langle #1 | #2 \rangle}

\newcommand{\Tr}{\operatorname{Tr}}
\newcommand{\avg}[1]{\left\langle#1\right\rangle}

\newcommand{\QAO}{\textsc{qao}\xspace}
\newcommand{\QA}{\textsc{qa}\xspace}
\newcommand{\SA}{\textsc{sa}\xspace}
\newcommand{\SQA}{\textsc{sqa}\xspace}
\newcommand{\QMC}{\textsc{qmc}\xspace}

\newcommand{\gmin}{g_\mathrm{min}}

\preprint{Preprint}

\begin{document}

\title{Quantum Monte Carlo Simulations of Tunneling in Quantum Adiabatic Optimization}

\author{Lucas~T.~Brady}
\affiliation{{Department of Physics, University of California, Santa Barbara, CA 93106-5110, USA}}
\author{Wim~van~Dam}
\affiliation{{Department of Computer Science, Department of Physics, University of California, Santa Barbara, CA 93106-5110, USA}}

\date{\today}
\begin{abstract}
We explore to what extent path-integral quantum Monte Carlo methods can
efficiently simulate the tunneling behavior of quantum adiabatic optimization
algorithms. Specifically we look at symmetric cost functions defined over $n$
bits with a single potential barrier that a successful optimization algorithm
will have to tunnel through. The height and width of this barrier depend on $n$,
and by tuning these dependencies, we can make the optimization algorithm succeed
or fail in polynomial time. In this article we compare the strength of quantum
adiabatic tunneling with that of path-integral quantum Monte Carlo methods. We
find numerical evidence that quantum Monte Carlo algorithms will succeed in the
same regimes where quantum adiabatic optimization succeeds.
\end{abstract}

\maketitle

\section{Introduction}
\label{sec:intro}
\subsection{Background}
Quantum Adiabatic Optimization (\QAO), first proposed by Farhi et al.\
\cite{Farhi2000}, is a quantum algorithm for determining the minimum of a cost
function by slowly evolving a Hamiltonian from one with known ground state to
one that has as its ground state the solution to an optimization problem. \QAO
relies on the quantum adiabatic theorem (see Jansen et al.\ \cite{Jansen} for a
proof), which roughly says that a system is guaranteed to stay in its ground state if
the Hamiltonian evolution time-scales are much larger than the square of the
inverse spectral gap. It was shown by, for example, Reichardt \cite{Reichardt} 
and Farhi et al.\ \cite{Farhi2002} that this algorithm might provide an 
exponential speed-up over naive local search algorithms. 
On the other hand, Farhi et al.\ \cite{Farhi2008} 
also described cases where it is no better than the quadratic speed up of 
Grover's quantum search.

\QAO is sometimes referred to as quantum annealing (\QA) in relation to
classical simulated annealing (\SA) where a simulated system's temperature is
slowly lowered reducing the probability of energetically less favorable states
until the ground state is reached at zero temperature. In recent years, there
has been a push to compare these two methods, often by using simulated quantum
annealing (\SQA) \cite{Martonak}. \SQA uses a path integral expansion of the
partition function for the evolving system to create a $(d+1)$ dimensional
classical system on which Monte Carlo techniques can be used. Instead of varying
the temperature as in \SA, \SQA varies the Hamiltonian in the same way as \QAO.

This path-integral Quantum Monte Carlo (\QMC) algorithm has been used to compare
classical \SA and \QA. Heim et al.\ \cite{Heim} among others have shown that \QMC
methods outperform classical \SA in several cases. In other situations Battaglia et al.\ \cite{Battaglia} 
showed that \SA can perform better than \QMC. In addition, new techniques in \SQA
through \QMC continue to be developed and improved, such as by Farhi et al.\ \cite{Farhi2011}.

\SQA through \QMC captures much of the power of \QAO, and for some problems
these two methods show correlation in their success rates while at the same time
remaining uncorrelated from classical \SA \cite{Boxio}. However, Hastings has
recently \cite{Hastings} constructed several examples where \QAO will find the
ground state in polynomial time whereas \QMC methods will take exponential time.

\subsection{Central Problem}

To directly compare the strengths and weaknesses \QAO and \QMC methods, this article will
look at their respective efficiencies as the two methods tunnel through a potential barrier. The
specific problem consists of a symmetric cost function on $n$ bits where each
basis state $x\in\{0,1\}^n$ is weighted by its Hamming weight $|x|$ in combination with a 
potential barrier centered at
$|x|=n/4$. Barriers of this form have been partially considered in the
context of \QAO by Reichardt \cite{Reichardt} who found that \QAO would succeed
in time polynomial in $n$ 
if the height and width of the barrier are both $\Omega(n^{1/4})$.

A simplified problem with a barrier of width $1$ was analyzed by Farhi et al.\
\cite{Farhi2002}, comparing \QAO and classical \SA. There \QAO was found to
succeed in polynomial time while classical \SA could not. Crosson and Deng
\cite{Crosson} showed that the same thin barrier limit is a case where \QMC
methods and \QAO both succeed together.

Muthukrishnan et al.\ \cite{Muthukrishnan} analyzed a similar problem where
instead of a barrier in the Hamming Weight, they have a plateau. This problem
has a constant gap, but they showed that \QAO still outperforms \SA, though both
run in polynomial time. They also showed that a non-adiabatic approach to \QA
could outperform \QAO in this case with a constant gap.

Our current goal is to extend the comparison of \QAO and \QMC methods to the
case of a varying barrier size. Therefore, we seek to determine if the
correlation between the two continues for the full case where the width and
height of the barrier are both powers of the number of qubits $n$.

\subsection{Organization}

In Section~\ref{sec:cost}, we will setup the particular problem we are working
with, defining our symmetric Hamiltonian and its tunable parameters. In Section
\ref{sec:gap} we will examine the energy eigenvalues of this Hamiltonian. We
will focus on the spectral gap between the ground state and first excited state
and will primarily use numerical diagonalization. The size of this spectral gap
determines how slowly adiabatic evolution must go in order to stay in the ground
state.

Section~\ref{sec:QMC} will outline and develop on our Monte Carlo method. We
will go through the approximations and how those approximations effect our
simulations; additionally, we will discuss our choice of update rules. In
Section~\ref{sec:results}, we present the results of our Monte Carlo simulations
and compare the scaling behavior of these simulations to the scaling behavior of
the spectral gap from Section~\ref{sec:gap}. Finally in
Section~\ref{sec:conclusion}, we discuss the limitations of our Quantum Monte
Carlo algorithm and present several avenues for extension and generalization of
our work.

\section{Hamming Weight with a Barrier}
\label{sec:cost}

Our problem is one discussed by Reichardt \cite{Reichardt}, and a simplified
version of it was analyzed by Crosson and Deng \cite{Crosson}. We consider a
symmetric cost function $f(|x|) = |x|+ b(|x|)$, where $|x|$ is the Hamming
Weight of the length $n$ bit string $x$, and $b(z)$ is some perturbing function.
We will take $b(z)$ to be some barrier, centered around $z = n/4$, that has
width and height proportional to $n^\alpha$. For ease of computation, we will
use
\begin{equation}
      b(z) = \begin{cases}
            n^\alpha & \text{when~}\left(\frac{n}{4}-\frac{1}{2}c\,n^\alpha\right)<z<
                        \left(\frac{n}{4}+ \frac{1}{2}c\,n^\alpha\right)\\
            0 & \text{otherwise}
      \end{cases},
\end{equation}
where $c$ is an $n$ independent constant. From now on we will say that this
barrier has size $c\,n^\alpha$. The full cost function will have
a global minimum at $|x| = 0$ and a local minimum at $|x| =
\left\lfloor{\frac{n}{4}+ \frac{1}{2}c n^\alpha}\right\rfloor+1$.

We will encode this problem into a Hamiltonian on a Hilbert space of $n$ qubits:
\begin{equation}
      H_1 = \sum_{x\in\{0,1\}^n} f(|x|) \ket{x}\bra{x}.
\end{equation}

In \QAO, we slowly transition from a Hamiltonian with a known ground state into
one with a desired ground state such as $H_1$ (e.g. in this problem, we want to
find the $|x|=0$ state). The standard initial Hamiltonian is
\begin{align} 
      H_0 & = \sum_{i=1}^n (\mathsf{H}_0)_i & \text{with}&  
      & \mathsf{H}_0 & = \frac{1}{2}
      \left(\begin{array}{cc}1&-1\\-1&1\end{array}\right),
\end{align}
where $i$ sums over all $n$ qubits. The ground state of this Hamiltonian is a
uniform superposition over all $\ket{x}$ states. Therefore, the ground state is
initially a binomial probability distribution over $|x|$ with width
$\sim\sqrt{n}$ centered at $|x|=n/2$. In \QAO, we create the Hamiltonian
\begin{equation} \label{eq:Hs}
      H(s) = (1-s) H_0 +s H_1,
\end{equation}
where $s$ goes from $0$ to $1$. If we vary $s$ slowly enough, the adiabatic
theorem says that the system will remain in the ground state. Therefore, the
system will be forced to tunnel through the potential barrier in order to reach
the true final ground state with $|x| = 0$. As $s$ changes, the first two energy
eigenlevels remain distinct and have some spectral gap $g(s)$. If the minimum
gap over $s$ is $\min_{s\in[0,1]}g(s)=\gmin$, then adiabatic evolution is
guaranteed to keep the system in the ground state if it takes time
${\Omega}(\gmin^{-2})$.

\section{Exact Spectral Gap}
\label{sec:gap}
To determine the minimum spectral gap $\gmin$ we will numerically diagonalize
the Hamiltonians $H(s)$. Using the symmetry of the Hamiltonians we are able to
do this accurately in the same range of finite $n$ that our Quantum Monte Carlo
simulations access. As a result, we will be able to compare the \QMC run-times
directly to the $1/\gmin^2$ quantity, rather than having to rely on
extrapolations to large $n$ behavior.

\subsection{Symmetrized Hamiltonian}

In order to diagonalize the $H(s)$ of Eq.~\ref{eq:Hs} for sizable $n$, we rely
on the symmetric subspace of our system. For each Hamming weight $0\leq h \leq
n$ we have that $H(s)$ is degenerate in the $\binom{n}{h}$ dimensional subspace
spanned by the vectors $\{\ket{x}:|x|=h\}$. Hence we see that the spectrum of
$H(s)$ has at most $n+1$ distinct eigenvalues, which will simplify our numerical
calculations significantly. We rewrite the Hamiltonian as follows.
\begin{align}
      H_\mathrm{sym}(s) =& \sum_{h=0}^n
            \left[\frac{(1-s)}{2}n+sf(h)\right]\ket{h}\bra{h}\\\nonumber
      +&\sum_{h=0}^{n-1}
            \left[-\frac{(1-s)}{2}\sqrt{(h+1)(n-h)}\right]\ket{h}\bra{h+1}
                  \\\nonumber
      +&\sum_{h=0}^{n-1}
            \left[-\frac{(1-s)}{2}\sqrt{(h+1)(n-h)}\right]\ket{h+1}\bra{h}
\end{align}

The spectral gap is then found by diagonalizing the resulting $(n+1)\times(n+1)$
tridiagonal matrix. Incidentally, the Hilbert space of this symmetrized system
is identical to that of a single spin-$n/2$ particle. In that spin
context, the Hamiltonian describes applying a magnetic field to the particle
where the field starts as a uniform field in the $-\hat{x}$ direction and then
rotates to one in the $\hat{z}$ direction with certain momentum modes picked out
as more energetic.

Furthermore, the adiabatic theorem states that \QAO run-time depends on the
minimum spectral gap as $s$ evolves, $\gmin$, so we minimize the gap as a
function of $s$. We restrict ourselves to $n$ divisible by $4$ so that the
barrier is centered on an integer Hamming weight. Since the barrier width
increases in integer steps, we only consider $n$s such that the width has just
increased (i.e.\ $n$ such that $\lfloor 1+ cn^\alpha\rfloor>\lfloor
1+c(n-4)^\alpha\rfloor$). Finally we require the barrier to have width less than
$n/2$, preferably much less, so that the $s=0$ ground state does not
have a significant overlap with the region of the barrier.

\subsection{Numerical Results}

\begin{figure}
      \resizebox{0.5\textwidth}{!}{\input{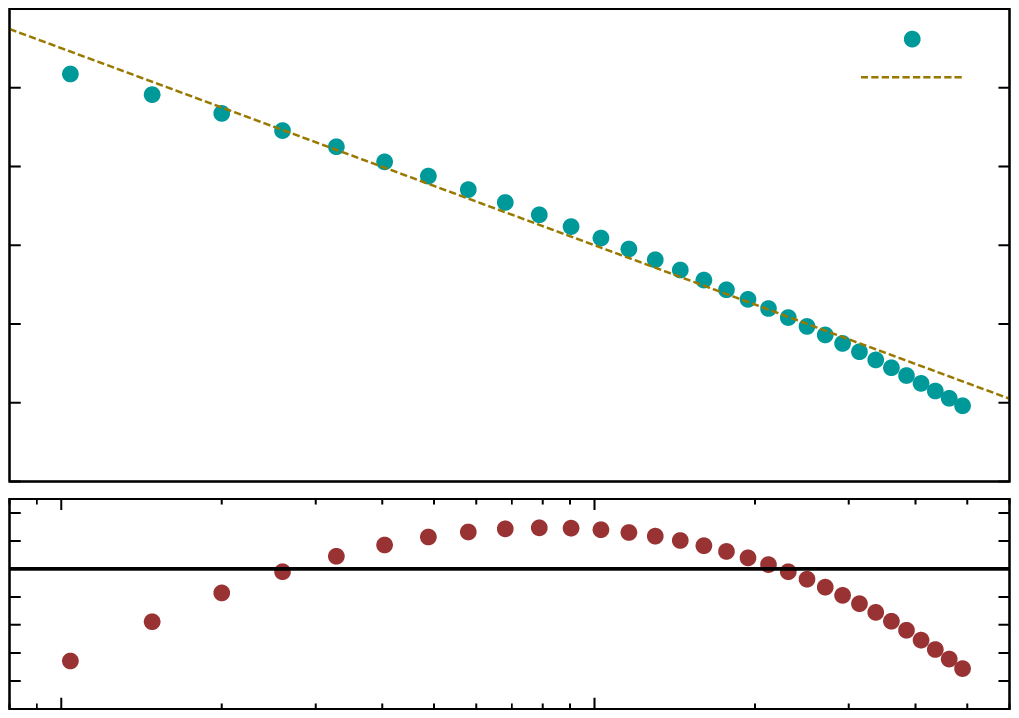}}
      \caption{
            {$\gmin$ vs.\ $n$ for barrier size $n^{0.5}$:}           
            We show a best fit linear regression through the $\log$-$\log$ data
            and plot the residuals of that linear fit versus the $\log$-$\log$
            data. The fact that the residuals curve down means that $\gmin$ is
            decreasing faster than a power law with $n$, indicating
            superpolynomial growth in the \QAO run-time
      }
      \label{fig:p=0.5_gap}
\end{figure}

\begin{figure}
      \resizebox{0.5\textwidth}{!}{\input{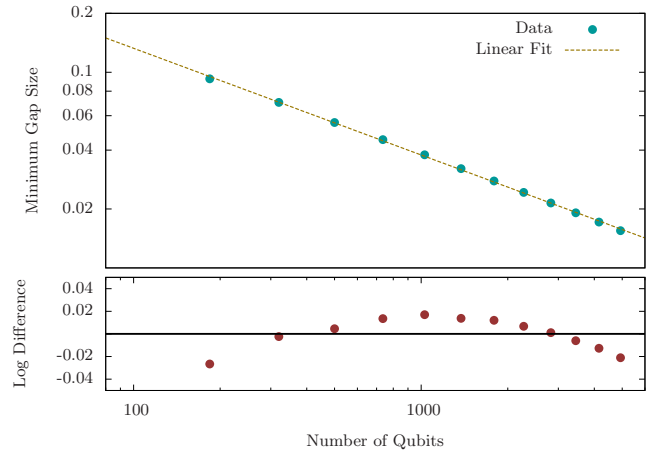}}
      \caption{
            {$\gmin$ vs.\ $n$  for barrier size  $n^{0.4}$:}
            The best fit linear regression to the $\log$-$\log$ data has
            residuals that curve downwards indicating superpolynomial growth in
            the \QAO run-time. Also, notice that $y$-axis scale on the residual
            plot is much smaller than in Fig.~\ref{fig:p=0.5_gap}, indicating
            that this scaling is not as strong as n the higher $\alpha$ case.
      }
      \label{fig:p=0.4_gap}
\end{figure}

\begin{figure}
      \resizebox{0.5\textwidth}{!}{\input{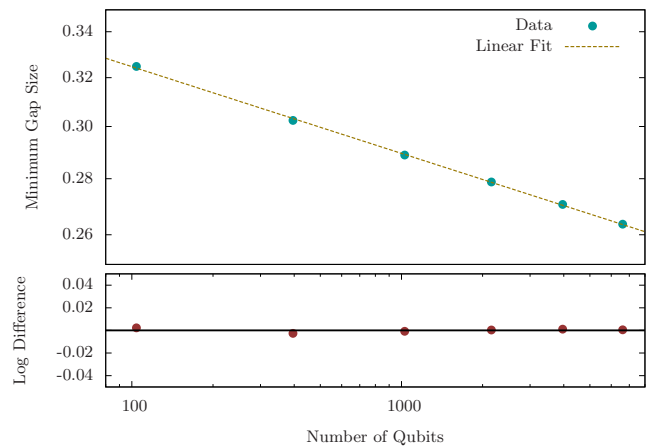}}
      \caption{
            {$\gmin$ vs.\ $n$  for barrier size  $n^{0.3}$:}
            The best fit linear regression to the $\log$-$\log$ data has
            residuals that are essentially zero, indicating polynomial scaling
            with $n$. We have used the same residual scale as in
            Fig.~\ref{fig:p=0.4_gap} to indicate just how small these residuals
            are. 
      }
      \label{fig:p=0.3_gap}
\end{figure}

In Fig.~\ref{fig:p=0.5_gap} we show the minimum gap for a barrier of size
$n^{0.5}$ as a function of $n$. The line drawn through the points is a linear
best fit to the $\log$-$\log$ data, and the plot below shows the residuals for
this fit. Since the residuals curve downward, the gap is decreasing faster than
a power law can account for; therefore, the running time for \QAO, which depends
on the gap $\gmin^{-2}$, is superpolynomial in $n$ for $\alpha = 0.5$. In
Figs.~\ref{fig:p=0.4_gap} and \ref{fig:p=0.3_gap}, we show similar plots for
$\alpha=0.4$ and $0.3$ respectively.

\begin{figure}
      \resizebox{0.5\textwidth}{!}{\input{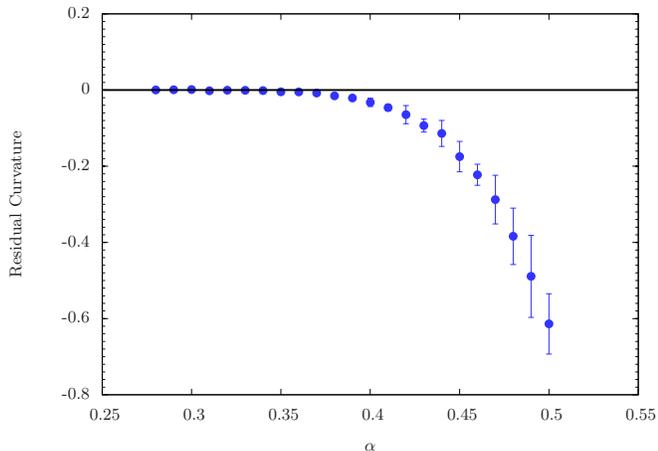}}
      \caption{
            {Deviation of $\gmin$ from Power Law in $n$:}
            For each barrier scaling power $\alpha$ at $c=1$ we found the
            spectral gap for $n$ between $100$ and $5000$ and tried to fit a
            linear curve to the $\log$-$\log$ plot of spectral gap versus $n$.
            What is displayed here is the curvature of the residuals from those
            fits. If the residuals are concave down (meaning negative curvature
            on this figure), the spectral gap is decreasing faster than a power
            law in $n$. Therefore, \QAO will become superpolynomial in $n$
            somewhere between $\alpha=0.33$ and $0.34$.
      }
      \label{fig:vary_gap}
\end{figure}

Varying $\alpha$, we do the same procedure, sweeping through a range of $n$ from
$100$ to $5000$ with $c=1$. The second derivative of these $\log$-$\log$
residuals can be used to estimate the curvature of those residuals (i.e. whether
they are concave up or down), and these second derivatives are plotted in
Fig.~\ref{fig:vary_gap}. The residual plots in Figs.~\ref{fig:p=0.5_gap},
\ref{fig:p=0.4_gap}, and \ref{fig:p=0.3_gap} are all used in the construction of
Fig.~\ref{fig:vary_gap}. Since the second derivative varies over the range of
$n$, the second derivative is averaged and the standard error is used as the
error bars. A negative number indicates superpolynomial running time, whereas
zero represents polynomial scaling.

The curvature in Fig.~\ref{fig:vary_gap} becomes negative by more than one error
bar starting at $\alpha=0.34$, which indicates that the quantum adiabatic
algorithm will undergo a transition from polynomial to exponential scaling
somewhere between $\alpha = 0.33$ and $0.34$.

It is a folklore result \cite{Goldstone} that as $n$ grows for a barrier with
height and width proportional to $n^\alpha$, the spectral gap decreases
asymptotically as:
\begin{equation}
      g_\mathrm{min} = 
      \begin{cases}
            \textrm{constant}                 & \text{if $\alpha<\frac{1}{4}$}\\
            1/\mathrm{polynomial}(n) & \text{if $\frac{1}{4} < \alpha< \frac{1}{3}$}\\
            1/\mathrm{exponential}(n) 
                              & \text{if $\frac{1}{3}< \alpha$}
      \end{cases}
      \label{eq:goldstone}
\end{equation}
Hence we expect a transition from polynomial to exponentially small gaps to
occur when $\alpha= 1/3$. This result meshes exactly with the results of our
simulations. Our numerical results are still useful in their own rights since
our \QMC calculations will be accessing finite $n$ values and it is important to
compare the \QMC results with equivalent gap results, and we need these results
to be aware of any possible small $n$ phenomena.

\begin{figure}
      \resizebox{0.5\textwidth}{!}{\input{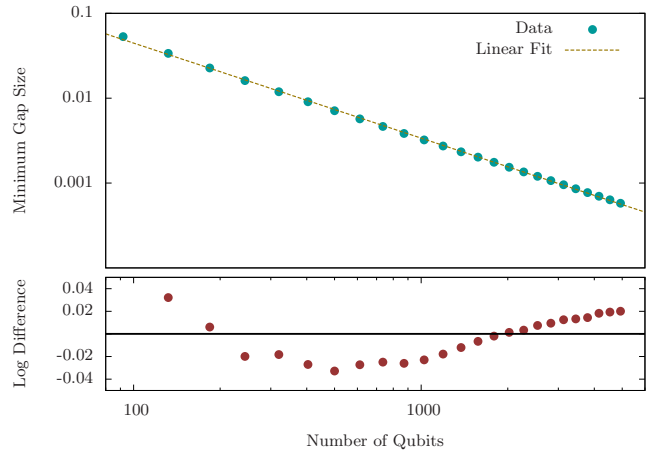}}
      \caption{
            {$\gmin$ vs.\ $n$  for barrier size  $2n^{0.4}$:}
            The best fit linear regression to the $\log$-$\log$ data has
            residuals that curve upwards indicating polynomial or subpolynomial
            decrease with $n$. At the end of the $n$ range, the residuals begin
            to curve down again, indicating the beginning of the superpolynomial
            region indicated by Figs.~\ref{fig:vary_gap} and
            \ref{fig:p=0.4_gap}.
      }
      \label{fig:p=0.4_gap_c=2}
\end{figure}

Additional numerical results indicate that the large $n$ scaling behavior in
Fig.~\ref{fig:vary_gap} does not hold for smaller $n$ when $c$ is large. For
instance Fig.~\ref{fig:p=0.4_gap} does display the large $n$ superpolynomial
behavior, but Fig.~\ref{fig:p=0.4_gap_c=2} does not. In
Fig.~\ref{fig:p=0.4_gap_c=2}, if we consider just the largest $n$, there are
indications that the residuals are becoming concave down at the end, indicating
that the superpolynomial scaling is starting at the end of the $n$ range we are
looking at.

The computational limits of our \QMC algorithm and computing facility mean that
some of the \QMC simulations in this article will be at lower $n$ where the
large $n$ scaling behavior is not yet dominant. In cases where we can access the
large $n$ scaling behavior, such as $\alpha=0.4$ and $c=1$ in
Fig.~\ref{fig:p=0.4_gap}, we will mention so in subsequent analysis. Largely, we
will be comparing \QMC running times with $\gmin^{-2}$ directly so that we can
see if \QMC running time scales polynomially with \QAO running time.

\section{Path-integral Quantum Monte Carlo}
\label{sec:QMC}
The path-integral \QMC algorithm \cite{Suzuki} is a method of simulating a
quantum mechanical system at finite inverse temperature $\beta$. The procedure
uses Trotter expansion to take an $n$ qubit quantum system to a classical system
of $n$ bits evolving in a discretized ``imaginary time'' dimension. These time
evolving states can then be treated as states in a Monte Carlo simulation that
samples possible paths of the system.

The Monte Carlo algorithm then picks paths with probability proportional to
their Boltzmann weights, so from these states, an expectation value for the
ground state energy can be obtained. We run the Monte Carlo algorithm for fixed
$s$ until we reach the ground state at that $s$ value and then transition to a
new $s$. This so called annealing schedule captures the same adiabaticity that
makes \QAO so powerful.

\subsection{Trotter Expansion}

To start, we take the partition function at finite inverse temperature $\beta$
and Trotter expand it into $T$ ``time''-slices
\begin{align}
      \label{eq:partition1}
      Z &= \Tr\left\{e^{-\beta H}\right\}\\\nonumber
      &= \lim_{T\to\infty}\sum_{x^{(0)},\dots, x^{(T-1)}}\left[\prod_{\tau=0}^{T-1}
            \bra{x^{(\tau)}}e^{-\frac{\beta}{T}H}\ket{x^{(\tau+1)}}\right],
\end{align}
where the sums go over each $x^{(\tau)}\in\{0,1\}^n$. In order to be in the
ground state, the temperature needs to be low, which means high $\beta$, but $T$
also needs to be much greater than $\beta$ in order for the Trotter
approximation to work well. In practice, we will take $\beta=32$ and $T\propto
n$ for reasons that will be discussed in subsection \ref{ssec:exp_app}. We also
have periodic boundary conditions $x^{(0)} = x^{(T)}$. The goal is to have the
operators act on these $\ket{x}$ basis states so that we can get a partition
function in terms of $c$-numbers. Each of the $T$ bases corresponds to a
different imaginary ``time'' slice of the system, so we are transforming our $n$
qubit system into an $n\times T$ lattice of classical bits with interactions
between adjacent time slices.

\subsection{Exponential Approximation}
\label{ssec:exp_app}

The Hamiltonian includes terms diagonal in the computational basis, which we
will call $H_d$, and off-diagonal terms, which we will call $H_o$. The goal is
to separate out these terms so that each operator can act on its own eigenbasis.
There are two approximations that can be used here: either a linear
approximation or an exponential approximation for $\beta/T\to 0$:
\begin{subequations}
\begin{align}
      \label{eq:approx1}
      e^{-\frac{\beta}{T}(H_d+H_o)} &= 1-\frac{\beta}{T}(H_d+H_o)
            +{O}((\beta/T)^2)\\
      \label{eq:approx2}
      e^{-\frac{\beta}{T}(H_d+H_o)} &= 
            e^{-\frac{\beta}{T}H_d}e^{-\frac{\beta}{T}H_o}
            +{O}((\beta/T)^2).
\end{align}
\end{subequations}
To first order these are both the same, but the additional terms in the
exponential change the algorithm significantly. The linear approximation only
includes one copy of the off-diagonal Hamiltonian, so adjacent Trotter time
slices would differ by at most a single bit. Single bit flips between adjacent
sites lend a nice sense of continuity to the time dimension, but they
necessitate larger $T$. The off-diagonal part of the Hamiltonian manifests
itself in the simulation as bit flips between adjacent time-slices, so in order
to get enough bit flips in the linear approximation, $T$ must be larger, whereas
the exponential approximation, with multiple adjacent bit flips, can be more
compact.

In Fye \cite{Fye}, there is a discussion of these two approximation methods where
they find that for local Hamiltonians the exponential approximation is more
robust and can be used with an $n$-independent $T$. The linear approximation
requires $T$ to increase with increasing $n$, making it less desirable. Our
Hamiltonian relies on the Hamming Weight, which is a non-local quantity, so
these results do not hold perfectly. We found that the exponential approximation
did require $T$ to have some dependence on $n$; however, numerically, we found
that dependence to be much smaller than the dependence of the linear
approximation. Therefore, we use the exponential approximation in this article.

Eventually, we will want to interpret the product of these exponentials as a
Boltzmann factor or probability for the given $n\times T$ configuration of the
system. In order to do this, the Boltzmann factors must be positive. In order to
ensure that our approximated exponentials remain positive, the Hamiltonian must
be one with ``no sign problem.'' This means that all the off-diagonal terms in
the Hamiltonian must be non-positive. To see why, consider Eq.~\ref{eq:approx1};
if the off-diagonal Hamiltonian contained negative terms, then this operator
would lead to negative terms if it were between non-identical states. This same
logic is true in Eq.~\ref{eq:approx2}. Our Hamiltonian has no sign problem, so
we are free to use these methods.

\subsection{Final Partition Function}

For an in depth derivation of the partition function see Appendix
\ref{app:partition}. Here, we will just cite the resulting partition function

\begin{align}
      \label{eq:partition2}
      Z &= \lim_{T\to\infty}\sum_{x^{(0)},\ldots, x^{(T-1)}}
            \left[\prod_{\tau=0}^{T-1}e^{-\frac{\beta}{T}
            \left((1-s)\frac{n}{2}+sf(|x^{(\tau)}|)\right)}\right.\\\nonumber
      &\times\left.\prod_{d=1}^{n}\left(
            e^{\frac{\beta}{T}\frac{(1-s)}{2}}
            +(-1)^{x_d^{(\tau)}-x_d^{(\tau+1)}}
            e^{-\frac{\beta}{T}\frac{(1-s)}{2}}\right)
            \right].
\end{align}

The first summation can be thought of as a sum over possible states, where a
state is a full configuration of the $n\times T$ bit lattice. The expression in
the square brackets is the Boltzmann factor for that configuration. The
Boltzmann factors are the unnormalized probabilities for the states, so they can
be used in a Metropolis algorithm to create a Monte Carlo simulation. The
Quantum Monte Carlo method consists of performing standard Monte Carlo methods
on this classical partition function which can then be used to gain information
about the original quantum system (e.g. see Appendix \ref{app:energy} for how to
extract the energy from this Monte Carlo simulation).

\subsection{Update Rules}

We follow the same update rule as Crosson and Deng \cite{Crosson}, where we
sweep through these $n\times T$ bits. One sweep consists of $n\times T$ updates,
where we go through each bit in the lattice separately. For that bit we try
flipping its value, and then compare the Boltzmann weight of the lattice before
and after the bit-flip. The acceptance rate of this bit flip is then equal to
the ratios of the Boltzmann factors before and after the flip. Once the sweep
has gone through every bit in the lattice, the sweep ends, and the algorithm
calculates the current ground state energy of the entire lattice based on the
results of Appendix \ref{app:energy}.

For the annealing schedule, we have a fixed $\Delta s = \frac{1}{100}$ and
change how much time we spend on each $s$ value. The algorithm calculates the
quantum mechanical energy (see Appendix \ref{app:energy}) of the system after
each sweep and moves onto the next $s$ value when the energy gets close enough
to the true ground state energy. This annealing schedule does use information
that the \QMC algorithm would not have in a normal simulation (namely the ground
state energy and spectral gap), but since our goal is to judge how long it takes
to reach the ground state rather than how long it takes the algorithm to realize
it has reached the ground state, this is appropriate.

The algorithm judges it is close enough to the true ground state when the
average energy over the last $100$ sweeps, $\avg{E(s)}_{100}$, is within $0.4$
spectral gaps, $g(s)$, of the true ground state energy, $E_{GS}(s)$:
\begin{equation}
\frac{\left|\avg{E(s)}_{100}-E_{GS}(s)\right|}{g(s)}<0.4
\end{equation}
In subsequent graphs, we will report the number of sweeps for each $s$ value. If
the algorithm has already satisfied this condition after the first $100$ steps,
it extrapolates back to when it first met the update condition and report that
as the number of sweeps.

\begin{figure}
      \resizebox{0.5\textwidth}{!}{\input{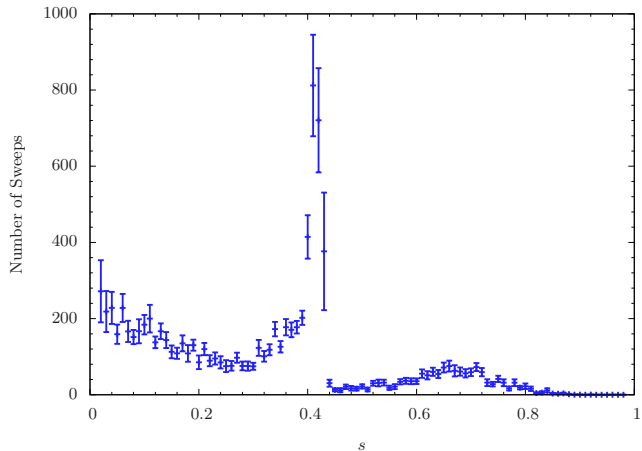}}
      \caption{
            {\QMC Sweeps vs.\ $s$ for barrier size $3 n^{0.5}$ at $n=116$:}
            This is averaged over $30$ simulations. The spike corresponds to
            tunneling through the potential barrier and is roughly at the
            location of the minimum spectral gap.
      }
      \label{fig:runtime_s_0.5}
\end{figure}

In Fig.~\ref{fig:runtime_s_0.5}, the results are shown for simulations using a
barrier of size $3n^{0.5}$, and $n=116$. Notice the spike in
run-time corresponding to tunneling through the potential barrier. The location
of this spike in $s$ corresponds to the location of the minimum spectral gap in
the exact problem, and this $s$ location always occurs in roughly the same
location across multiple values of $n$ and $\alpha$. In the next
section when we report the run-time of the \QMC simulations, we will report the
total number of sweeps taken between $s=0.3$ and $s=0.5$. For all of our
simulations, this $s$ range captures the run-time spike and some of the
surrounding area while ignoring any low $s$ initialization artifacts or high $s$
tailing-off.

\section{Numerical Monte Carlo Results}
\label{sec:results}

In this section, we will explore a few different values of the barrier scaling
power, $\alpha$, and the width scaling coefficient, $c$, using the \QMC methods
developed in the previous section. For most of the simulations considered here
our number of Trotter slices is related to the number of qubits through $T=4n$.
In reporting \QMC times, we will report the number of sweeps each simulation
took while going through the critical $s$ region. There are $n\cdot T$
Metropolis steps per sweep, so the actual run-time of the algorithm depends
polynomially on the number of sweeps.

\begin{figure}
      \resizebox{0.5\textwidth}{!}{\input{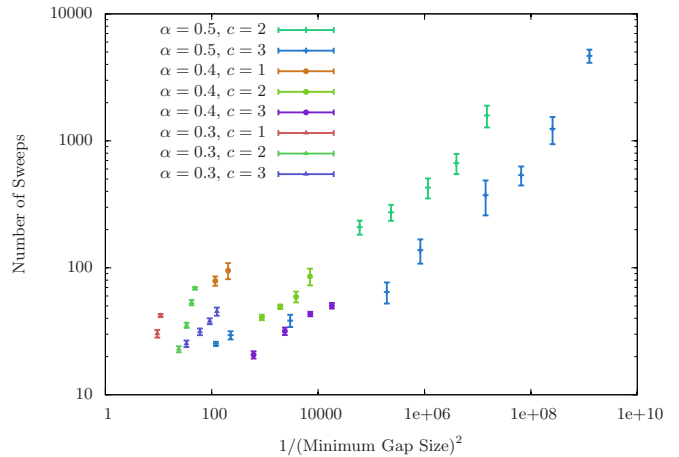}}
      \caption{
            {\QMC Sweeps vs.\ $\gmin$:}
            This lists all our data together; a further breakdown of this data
            is available in Figs.~\ref{fig:p=0.5_gap_vs_QMC},
            \ref{fig:p=0.4_gap_vs_QMC}, and \ref{fig:p=0.3_gap_vs_QMC}. Notice
            that there is an obvious strong correlation between required and
            sufficient \QMC sweeps and the gap. More analysis, specific to the
            different $\alpha$ values can be found in
            Figs.~\ref{fig:p=0.5_gap_vs_QMC}, \ref{fig:p=0.4_gap_vs_QMC}.
      }
      \label{fig:full_gap_vs_QMC}
\end{figure}

In Fig.~\ref{fig:full_gap_vs_QMC}, we show the full results of our \QMC
simulations, comparing the run-times of these algorithms to the corresponding
$\gmin^{-2}$. There is a strong correlation between these two quantities, which
at least indicates some relation. The following sections will breakdown this
data by $\alpha$ value and analyze it independently.

\subsection{Barriers Proportional to $\mathbf{\emph{n}^{0.5}}$}

To start, we will focus on $\alpha=0.5$. Based on Fig.~\ref{fig:vary_gap}, this
size of barrier has \QAO run-times that scale superpolynomially with $n$.
Practically, we are able to run \QMC simulations with $n$ ranging up to $\sim
220$ qubits. For this regime of $n$, small $n$ effects mask the superpolynomial 
scaling of the gap for $c=3$ but not for $c=2$.

Note that $c=2$ leads to smaller spectral gaps than $c=3$ at fixed $n$. From
trial and error, we found that the smaller gap sizes mean that the Trotter
approximation needs to be better in order to get sensible results. Thus, for
$c=2$, $T=16n$ rather than the usual $T=4n$. This necessity to improve the \QMC
for simulations with smaller gap sizes lends significant credence to the idea
that the \QMC algorithm depends heavily on the spectral gap itself.

\begin{figure}
      \resizebox{0.5\textwidth}{!}{\input{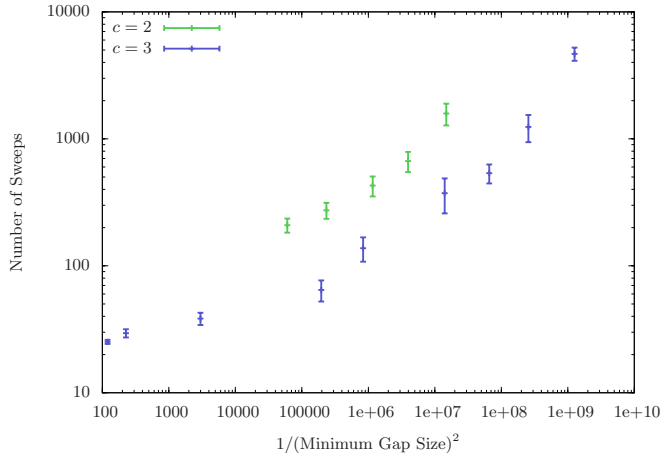}}
      \caption{
            {\QMC Sweeps vs.\ $\gmin$ for barrier size $cn^{0.5}$:}
            The number of sweeps is increasing faster than a power law with the
            inverse gap size, indicating the our specific \QMC algorithm is
            worse than \QAO in this case. For $c=2$ (green), $n$ ranges from
            $84$ to $172$, and for $c=3$ (blue), $n$ ranges from $88$ to $216$.
      }
      \label{fig:p=0.5_gap_vs_QMC}
\end{figure}

The \QMC run-time (averaged over multiple simulations) as a function of
$\gmin^{-2}$ is shown in Fig.~\ref{fig:p=0.5_gap_vs_QMC} Notice that the data in
this figure does not lie along a straight line, so the \QMC run-times seem to be
increasing at a rate faster than polynomially in the inverse gap. This lack of a
power law could be caused by three possible effects.

It is possible that this means the \QMC algorithm does indeed scale
superpolynomially with $\gmin^{-2}$. An alternative is that this curvature is
due to small $n$ effects that are still prevalent even for $n$ in the several
hundreds. Especially for $c=3$ and lower $n$, there is overlap between the
initial $s=0$ ground state distribution and the barrier, which could account for
the apparent deviation from a power law here. Additionally, this curvature could
be an indication of deficiencies in our \QMC implementation specifically. As
will be discussed in the next section, our algorithm has some notable
approximations and simplifications that could be leading to this discrepancy.

\subsection{Barriers Proportional to $\mathbf{\emph{n}^{0.4}}$}

\begin{figure}
      \resizebox{0.5\textwidth}{!}{\input{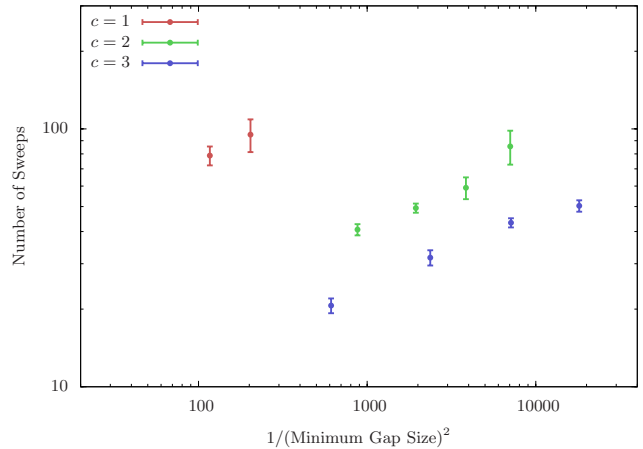}}
      \caption{
            {\QMC Sweeps vs.\ $\gmin$ for barrier size $cn^{0.4}$:}
            There appears to be a linear relationship here, indicating that \QMC
            performance and \QAO performance are polynomially related in this
            region. For $c=1$ (red), $n$ ranges from $184$ to $320$, for $c=2$
            (green), $n$ ranges from $132$ to $320$, and for $c=3$ (blue), $n$
            ranges from $116$ to $224$.
      }
\label{fig:p=0.4_gap_vs_QMC}
\end{figure}

For $\alpha = 0.4$, the \QMC simulations are able to go up to $\sim 320$ qubits.
In this regime of $n$, small $n$ effects mean that the gap is not
superpolynomial for $c=3,2$ (see Fig.~\ref{fig:p=0.4_gap_c=2}) but it is for
$c=1$ (see Fig.~\ref{fig:p=0.4_gap}). In Fig.~\ref{fig:p=0.4_gap_vs_QMC} we have
compared the \QMC run-times directly to the spectral gap. Notice that in this
case, there does seem to be a linear relationship between the $\log$-$\log$
data. Many of the deficiencies in our specific implementation are less
pronounced in this case than in the $\alpha=0.5$ case since the barrier is
smaller. There is less overlap between the initial ground state and the barrier,
which could also mean these simulations suffer less from small $n$ effects than
the $\alpha=0.5$ simulations.

\begin{figure}
      \resizebox{0.5\textwidth}{!}{\input{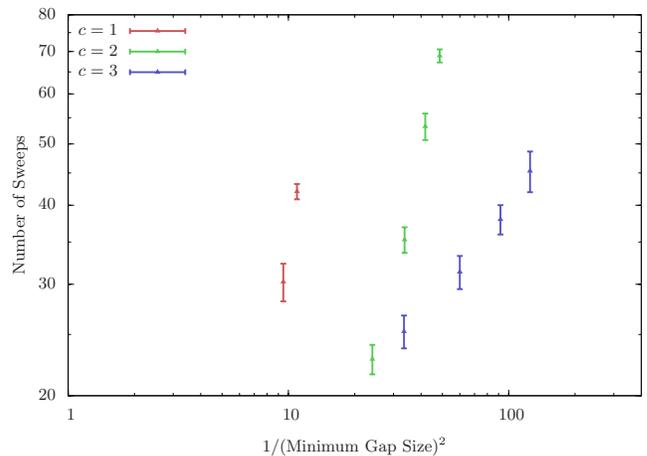}}
      \caption{
            {\QMC Sweeps vs.\ $\gmin$ for barrier size $cn^{0.3}$:}
            There appears to be a linear relationship here, indicating that \QMC
            performance and \QAO performance are polynomially related in this
            region. For $c=1$ (red), $n$ ranges from $104$ to $396$, for $c=2$
            (green), $n$ ranges from $104$ to $660$, and for $c=3$ (blue), $n$
            ranges from $104$ to $396$.
            }
\label{fig:p=0.3_gap_vs_QMC}
\end{figure}

\subsection{Barriers Proportional to $\mathbf{\emph{n}^{0.3}}$}

Finally for $\alpha=0.3$, numerical diagonalization indicates the gap decreases
polynomially in $n$ for low and high $n$, no matter what $c$ is chosen. Since
the width of the barrier does not increase often for such a low scaling power
$\alpha$, the number of $n$ accessible to the \QMC simulations is low here. Our
data is displayed in Fig.~\ref{fig:p=0.3_gap_vs_QMC}. Notice that there does
seem to be a linear relationship on the $\log$-$\log$ scale between inverse gap
size and run-time here, though it is partially masked by the dearth of data
points. However, this does seem to indicate a polynomial relationship between
\QMC run-time and $\gmin^{-2}$.

\begin{figure}
      \resizebox{0.5\textwidth}{!}{\input{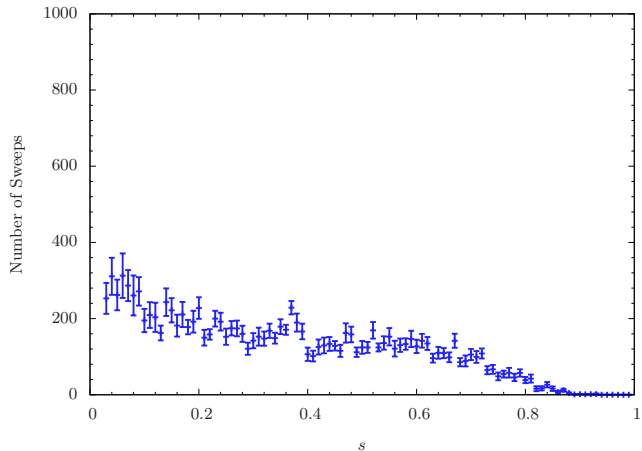}}
      \caption{
            {\QMC Sweeps vs.\ $s$ for barrier size $3 n^{0.3}$ at $n=116$:}
            This is averaged over $30$ simulations. Notice that unlike
            Fig.~\ref{fig:runtime_s_0.5}, there is no noticeable spike here
            corresponding to tunneling.
      }
      \label{fig:runtime_s_0.3}
\end{figure}

Additionally, a plot of run-time versus $s$ for higher powers, such as in
Fig.~\ref{fig:runtime_s_0.5}, shows a noticeable spike right at the tunneling
location. For lower powers, such as $\alpha=0.3$ as shown in
Fig.~\ref{fig:runtime_s_0.3}, there is no noticeable tunneling spike in the
run-time. From our simulation results, it seems that the distinction between
spikes and no spikes corresponds with the superpolynomial scaling cutoff we saw
in the spectral gap in Section~\ref{sec:gap}.

\section{Conclusion}
\label{sec:conclusion}

First, in Section~\ref{sec:gap}, we numerically verified a folklore
result\cite{Goldstone} about the relationship between $n$ and the minimum gap
$\gmin$. We showed that $\gmin$ scales polynomially with $n$ for barriers whose
height and width grow like $\alpha<\frac{1}{3}$ but that for
$\alpha>\frac{1}{3}$, the minimum gap decreases faster than a power law. This
indicates that \QAO can succeed in finding the true ground state in polynomial
time only for $\alpha<\frac{1}{3}$.

Our numerical results with Quantum Monte Carlo simulations show that above
$\alpha=\frac{1}{3}$, there is a clear slowdown in the \QMC algorithm (see
Fig.~\ref{fig:runtime_s_0.5}) whose location in $s$ corresponds well with the
location of the minimum gap in \QAO. This slowdown all but disappears for lower
$\alpha$ (see Fig.~\ref{fig:runtime_s_0.3}) where the \QMC algorithm has little
trouble tunneling through the potential barrier.  This is strong evidence that
there is a correlation between spectral gap and \QMC performance.

Furthermore, in Section~\ref{sec:results}, we showed that there is indeed a
correlation between gap size and \QMC run-time. For $\alpha$s less than
$\frac{1}{3}$, we see data consistent with a polynomial relationship between
\QMC run-time and $\gmin^{-2}$. This relationship is more difficult to discern
for $\alpha>\frac{1}{3}$ with there seeming to be either a polynomial or
superpolynomial relationship. The lack of a solid polynomial relationship could
be due to small $n$ effects which are more prevalent in our simulations for
higher $\alpha$, or it could also be due to inadequacies in our \QMC
implementation rather than \QMC algorithms in general

Most notably our algorithm keeps a fixed $\Delta s$ throughout its annealing
schedule and relies on spending more time on each $s$ value rather than
decreasing the size of the $s$ step. A more advanced algorithm could also
dynamically update $s$ to move more slowly through problem regions.

For the most part, our simulations also keep the number of Trotter time steps
$T=4n$. While $T=4n$ is sufficient for the region of parameter space discussed in
this article, it is possible that other Trotterization divisions would be more
efficient

Of course our work can also be extended by considering different regions in
parameter space of the Hamiltonian. The scaling of the height and width are
varied together using $\alpha$ in our analysis, but they can be varied
independently. Additionally, the shape of the barrier can be made more
complicated than the simple step used here. More generally, this procedure of
applying \QMC algorithms with annealing schedules can be used with other
Hamiltonians to gain insight into the relationship between \QAO and classical
computing.

\acknowledgements
This material is based upon work supported by the National Science Foundation
      under Grant No.~1314969.
We also acknowledge support from the Center for Scientific Computing from the
      CNSI, MRL: an NSF MRSEC (DMR-1121053) and NSF CNS-0960316.
We thank Aram Harrow for useful conversations.


\appendix

\section{Derivation of Partition Function}
\label{app:partition}

We will start with Eq.~\ref{eq:partition1}, and our goal will be to derive Eq.
\ref{eq:partition2} as well as an estimator for our quantum mechanical ground
state energy. Our first step will involve inserting our exponential
approximation scheme so that (we will start using hats on operators to avoid
confusion)
\begin{align}
      Z &= \lim_{T\to\infty} \sum_{x^{(0)},\ldots, x^{(T-1)}}\left[\prod_{\tau=0}^{T-1}\bra{x^{(\tau)}}
            e^{-\frac{\beta}{T}\hat{H}_d}
            e^{-\frac{\beta}{T}\hat{H}_o}\ket{x^{(\tau+1)}}\right],
\end{align}
where the sums go over each $x^{(\tau)}\in\{0,1\}^n$.

Here $\hat{H}_o$ and $\hat{H}_d$ are the off-diagonal and diagonal parts of the
Hamiltonian, given by
\begin{align}\nonumber
      \hat{H}_d &\equiv \sum_{x\in\{0,1\}^n}
            \left[(1-s)\frac{n}{2}+sf(|x|)\right]\ket{x}\bra{x}\\\nonumber
      \hat{H}_o &\equiv \sum_{\langle x,y\rangle}
            \left[-\frac{(1-s)}{2}\right]\ket{x}\bra{y}.
\end{align}

The sum in $\hat{H}_o$ is over nearest neighbor sites (i.e.\ bit strings $x$ and
$y$ that differ by one bit flip). Since $\hat{H}_d$ is diagonal in the
computational basis, we can just have it act on our basis states pulling out the
eigenvalues $H_d(x) = (1-s)\frac{n}{2}+sf(|x|)$.  

\begin{align}
      Z &= \lim_{T\to\infty}\sum_{x^{(0)},\ldots, x^{(T-1)}}
            \left[\prod_{\tau=0}^{T-1}e^{-\frac{\beta}{T}H_d(x^{(\tau)})}\right]
            \\\nonumber
      &\times\left[\prod_{\tau=0}^{T-1}
            \bra{x^{(\tau)}}e^{-\frac{\beta}{T}\hat{H}_o}\ket{x^{(\tau+1)}}
            \right].
\end{align}

Next, we will claim that there is an orthonormal basis $\ket{k^{(\tau)}}$ that is
the eigenbasis for $\hat{H}_o$, whose eigenvalues are $H_o(k^{(\tau)})$. We can
insert a complete set of these states at ever time slice to get
\begin{align}
      Z &=\lim_{T\to\infty} 
            \left[\sum_{x^{(0)},\ldots, x^{(T-1)}}\sum_{k^{(0)},\ldots, k^{(T-1)}}\right]
            \\\nonumber
      &~~~\left[\prod_{\tau=0}^{T-1}e^{-\frac{\beta}{T}H_d(x^{(\tau)})}
            e^{-\frac{\beta}{T}H_o(k^{(\tau)})}\braket{x^{(\tau)}}{k^{(\tau)}}
            \braket{k^{(\tau)}}{x^{(\tau+1)}}\right].
\end{align}

To find these $\ket{k}$ states, we just need to diagonalize $\hat{H}_o$. This
operator can be represented by a translationally invariant matrix on an $n$
dimensional hypercubic lattice (where each dimension is two sites long) with
periodic boundary conditions and nearest neighbor interactions. These properties
mean that the eigenstates of $\hat{H}_o$ are simply the Brillouin Zone lattice
sites. If we represent each Brillouin Zone lattice site using $k\in\{0,1\}^n$,
then these lattice sites can be represented in the $\ket{x}$ basis by
\begin{equation}
      \ket{k} = \sum_{x\in\{0,1\}^n}e^{i\pi\vec{k}\cdot\vec{x}}\ket{x}.
\end{equation}

Using standard Brillouin Zone methods for translationally invariant matrices, we
can work out that the eigenvalues of our off-diagonal Hamiltonian are
\begin{equation}
      H_o(k) = -\frac{(1-s)}{2}\sum_{d=1}^n(1-2k_d).
      \label{eq:H_o}
\end{equation}
Furthermore, the overlap between $\ket{x}$ and $\ket{k}$ states is given by
\begin{equation}
      \braket{x}{k} = (-1)^{\vec{k}\cdot\vec{x}}.
      \label{eq:xkinner}
\end{equation}

Inserting Eqs.~\ref{eq:xkinner} and \ref{eq:H_o} back into our partition function gives us
\begin{align}
      Z &=\lim_{T\to\infty} \sum_{x^{(0)},\ldots, x^{(T-1)}}
            \left[\prod_{\tau=0}^{T-1}
            e^{-\frac{\beta}{T}H_d(x^{(\tau)})}\right]\\\nonumber
      &\times
            \left[\prod_{\tau=0}^{T-1}\sum_{k^{(\tau)}}\prod_{d=1}^n
            e^{\frac{\beta}{T}\frac{(1-s)}{2}(1-2k_d^{(\tau)})}
            (-1)^{k^{(\tau)}_d(x^{(\tau)}_d-x^{(\tau+1)}_d)}
            \right].
\end{align}
We can rewrite $\sum_{k^{(\tau)}}\prod_{d=1}^n \to
\prod_{d=1}^n\sum_{k^{(\tau)}_d=0,1}$. Focusing on just the important part and
dropping the $\tau$ labels in favor of labeling the two bit strings by $x$ and $y$,
we get
\begin{align}
      \prod_{d=1}^{n}\sum_{k_d=0,1}
            e^{\frac{\beta}{T}\frac{(1-s)}{2}(1-2k_d)}
            (-1)^{k_d(x_d-y_d)} \\\nonumber
      = \prod_{d=1}^{n}\left[
            e^{\frac{\beta}{T}\frac{(1-s)}{2}}
            +(-1)^{x_d-y_d}
            e^{-\frac{\beta}{T}\frac{(1-s)}{2}}\right].
\end{align}
Note that we have now eliminated the $k$ variables entirely.
Inserting this simplification lets us exactly recover Eq.~\ref{eq:partition2}:
\begin{align}
      Z &=\lim_{T\to\infty} \sum_{x^{(0)},\ldots, x^{(T-1)}}
            \left[\prod_{\tau=0}^{T-1}
            e^{-\frac{\beta}{T}\left((1-s)\frac{n}{2}+sf(|x^{(\tau)}|)\right)}
            \right.\\\nonumber
      &\times\left.\prod_{d=1}^{n}\left(
            e^{\frac{\beta}{T}\frac{(1-s)}{2}}
            +(-1)^{x_d^{(\tau)}-x_d^{(\tau+1)}}
            e^{-\frac{\beta}{T}\frac{(1-s)}{2}}\right)
            \right].
\end{align}

\section{Derivation of Energy Estimators}
\label{app:energy}
Next, we need to look at what the expectation value of a quantum operator is in
the Trotter expanded formalism.  By definition, we have
\begin{equation}
      \avg{\hat{O}} = \frac{1}{Z}\Tr\left\{\hat{O}e^{-\beta \hat{H}}\right\}.
\end{equation}

When we do the Trotter expansion we do not and should not expand $\hat{O}$ as we
do the exponential. In fact after the Trotter expansion, we will still only have
one copy of $\hat{O}$ still, so the original copy of $\hat{O}$ will just be with
one of the time slices. For convenience, we will put it with the very first time
slice, so that after Trotterization, we are looking at:

\begin{align}
      \avg{\hat{O}} &=\lim_{T\to\infty} \frac{1}{Z}
            \sum_{x^{(0)},\ldots, x^{(T-1)}}
            \bra{x^{(T-1)}}e^{-\frac{\beta}{T}\hat{H}_d}
            e^{-\frac{\beta}{T}\hat{H}_o}\hat{O}\ket{x^{(0)}}
            \nonumber\\\nonumber
      &\times\left[\prod_{\tau=0}^{T-2}
            \bra{x^{(\tau)}}e^{-\frac{\beta}{T}\hat{H}_d}
            e^{-\frac{\beta}{T}\hat{H}_o}\ket{x^{(\tau+1)}}\right],\\
      \avg{\hat{O}} &=\lim_{T\to\infty} \frac{1}{Z}
            \sum_{x^{(0)},\ldots, x^{(T-1)}}
            \frac{\bra{x^{(T-1)}}e^{-\frac{\beta}{T}\hat{H}_d}
            e^{-\frac{\beta}{T}\hat{H}_o}\hat{O}\ket{x^{(0)}}}{\bra{x^{(T-1)}}
            e^{-\frac{\beta}{T}\hat{H}_d}
            e^{-\frac{\beta}{T}\hat{H}_o}\ket{x^{(0)}}}
            \nonumber\\
      &\times\left[\prod_{\tau=0}^{T-1}\bra{x^{(\tau)}}
            e^{-\frac{\beta}{T}\hat{H}_d}
            e^{-\frac{\beta}{T}\hat{H}_o}
            \ket{x^{(\tau+1)}}\right],
\end{align}

Next consider the probability of obtaining a specific configuration, 
$\{x^{(\tau)}\}$, of our $n\times T$ lattice of bits:
\begin{equation}
      p\left(\left\{x^{(\tau)}\right\}\right) \equiv 
            \frac{1}{Z}\left[\prod_{\tau=0}^{T-1}
            \bra{x^{(\tau)}}e^{-\frac{\beta}{T}\hat{H}_d}
            e^{-\frac{\beta}{T}\hat{H}_o}\ket{x^{(\tau+1)}}\right]
      \label{eq:Boltz_prob}
\end{equation}
Using Eq.~\ref{eq:Boltz_prob}, the average becomes
\begin{align}
      \avg{\hat{O}} &=\lim_{T\to\infty}
            \sum_{x^{(0)},\ldots, x^{(T-1)}}p\left(\left\{x^{(\tau)}\right\}\right)\\\nonumber
      &\times\frac{\bra{x^{(T-1)}}e^{-\frac{\beta}{T}\hat{H}_d}
            e^{-\frac{\beta}{T}\hat{H}_o}\hat{O}\ket{x^{(0)}}}
            {\bra{x^{(T-1)}}e^{-\frac{\beta}{T}\hat{H}_d}
            e^{-\frac{\beta}{T}\hat{H}_o}\ket{x^{(0)}}} 
            .
\end{align}

The \QMC method will specifically use the average energy: $\avg{\hat{H}} =
\avg{\hat{H}_d}+\avg{\hat{H}_o}$. Starting with $\avg{\hat{H}_d}$, the operator
is already acting on its eigenstates, so the average becomes
\begin{equation}
      \avg{\hat{H}_d} = \lim_{T\to\infty}
            \sum_{x^{(0)},\ldots, x^{(T-1)}}
            \left[H_d(x^{(0)})p\left(\left\{x^{(\tau)}\right\}\right)\right].
\end{equation}

In actual simulations, the estimator $H_d(x^{(0)})\to
\frac{1}{T}\sum_{\tau=0}^{T-1}H_d(x^{(\tau)})$ is used so that information from
the entire time dimension can enter the statistics.

Moving onto $\avg{\hat{H}_o}$ and focusing on just the relevant piece we have
(replacing $x^{(T-1)}\to x$ and $x^{(0)}\to y$ for notational convenience):
\begin{equation*}
      \frac{\bra{x}e^{-\frac{\beta}{T}\hat{H}_d}
      e^{-\frac{\beta}{T}\hat{H}_o}\hat{H}_o\ket{y}}
      {\bra{x}e^{-\frac{\beta}{T}\hat{H}_d}
      e^{-\frac{\beta}{T}\hat{H}_o}\ket{y}},
\end{equation*}
we can insert $k$ resolutions of the the identity in the top and bottom to get
\begin{widetext}\begin{equation}
      \frac{
            \sum_{k\in\{0,1\}^n}
            e^{-\frac{\beta}{T}H_o(k)}
            H_o(k)
            \braket{x}{k}
            \braket{k}{y}
      }
      {
            \sum_{k'\in\{0,1\}^n}
            e^{-\frac{\beta}{T}H_o(k')}
            \braket{x}{k'}
            \braket{k'}{y}
      }=
      \frac{
            -\frac{(1-s)}{2}
            \sum_{k\in\{0,1\}^n}
            e^{\frac{\beta}{T}\frac{(1-s)}{2}\sum_{d=1}^n(1-2k_d)}
            \sum_{p=1}^n(1-2k_{p})
            (-1)^{k\cdot(x-y)}
      }
      {
            \sum_{k'\in\{0,1\}^n}
            e^{\frac{\beta}{T}\frac{(1-s)}{2}\sum_{d=1}^n(1-2k'_d)}
            (-1)^{k'\cdot(x-y)}
      }
\end{equation}\end{widetext}
Next, we pull out what we can and switch $\sum_{k\in\{0,1\}^n}\prod_{d=1}^n \to
\prod_{d=1}^n \sum_{k_d=0,1}$:

In a given $p$ element, the term in the product will be the same in the
numerator and denominator if $d\neq p$, so the terms in the product cancel
except in the case where $d=p$:

\begin{equation}
      -\frac{(1-s)}{2}\sum_{p=1}^n
      \frac{
            e^{\frac{\beta}{T}\frac{(1-s)}{2}}
            -
            (-1)^{(x_p-y_p)}
            e^{-\frac{\beta}{T}\frac{(1-s)}{2}}
      }
      {
            e^{\frac{\beta}{T}\frac{(1-s)}{2}}
            +
            (-1)^{(x_p-y_p)}
            e^{-\frac{\beta}{T}\frac{(1-s)}{2}}
      }
      \label{eq:exp_frac}
\end{equation}

Inserting Eq.~\ref{eq:exp_frac} into the off-diagonal energy estimator gives
\begin{align}
      \avg{\hat{H}_o}
            &=\lim_{T\to\infty}
            \sum_{x^{(0)},\ldots, x^{(T-1)}}
            p\left(\left\{x^{(\tau)}\right\}\right)\\\nonumber
      &\times\left[
            -\frac{(1-s)}{2}\sum_{p=1}^n
            \frac{
                  e^{\frac{\beta}{T}\frac{(1-s)}{2}}
                  -
                  (-1)^{(x_p^{(0)}-x_p^{(T-1)})}
                  e^{-\frac{\beta}{T}\frac{(1-s)}{2}}
            }
            {
                  e^{\frac{\beta}{T}\frac{(1-s)}{2}}
                  +
                  (-1)^{(x_p^{(0)}-x_p^{(T-1)})}
                  e^{-\frac{\beta}{T}\frac{(1-s)}{2}}
            }
            \right]
\end{align}

Again, we typically average over the result for the different time slices in the
actual simulation.

\end{document}